\documentclass{article}

\usepackage{arxiv}

\usepackage[utf8]{inputenc} 
\usepackage[T1]{fontenc}    
\usepackage{hyperref}       
\usepackage{url}            
\usepackage{booktabs}       
\usepackage{amsfonts}       
\usepackage{nicefrac}       
\usepackage{microtype}      
\usepackage{lipsum}
\usepackage{amsmath} 
\usepackage{graphicx}
\usepackage[numbers]{natbib}
\graphicspath{ {./images/} }

\title{DiffSWE2d: a differentiable Shallow Water Equations solver for
end-to-end flood and tsunami modelling}

\author{
 Zhonghou Xu \\
  Earth Sciences New Zealand\\
  Hamilton, 3216 \\
  \texttt{zhonghou.xu@earthsciences.nz} \\
}

\begin{document}
\maketitle
\begin{abstract}
Solving inverse and optimisation problems with traditional shallow water equations (SWE) solvers can be computationally expensive, particularly when gradients with respect to model inputs or parameters must be estimated through repeated forward simulations. In this paper, we introduce \textbf{DiffSWE2d}, an open-source differentiable shallow water equations solver for end-to-end flood and tsunami modelling implemented in PyTorch. By leveraging automatic differentiation, DiffSWE2d represents the time-marching physics as a differentiable computational graph, enabling gradients to be propagated directly through the numerical solver. We validate the solver against two established benchmark cases and demonstrate its application to tsunami waveform inversion, showing its ability to infer model inputs through gradient-based optimisation. DiffSWE2d provides a flexible framework for integrating physics-based hydrodynamic modelling with modern optimisation and machine learning methods. The source code and reproducible examples are publicly available at:
\url{https://github.com/ZhonghouXu/DiffSWE2d}
\end{abstract}


\section{Introduction}
\label{sec:1}
Shallow water equations (SWE) are widely used for flood and tsunami modelling because they provide an effective balance between physical fidelity and computational efficiency. With the rapid development of high-performance computing, particularly graphics processing units (GPUs), the computational efficiency of SWE solvers has improved substantially, enabling large-scale and high-resolution flood and tsunami simulations \citep{bosserelleBGFloodGPUAdaptive2021, maciasPerformanceBenchmarkingTsunamiHySEA2017, vacondioNonuniformEfficientGrid2017, xiaFullscaleFluvialFlood2019}. However, most traditional SWE solvers are primarily designed for forward modelling, in which model outputs are computed from prescribed input parameters through a time-marching numerical scheme. Many practical applications instead require inverse or optimisation problems, where model parameters or inputs must be inferred or optimised based on observations or desired model outputs. Examples include inferring tsunami waveforms, estimating bathymetry and bottom roughness, optimising topography for flood mitigation, and designing sensor networks. Such problems typically require repeated forward simulations and gradient estimation, making them computationally expensive even when the underlying forward solver is highly accelerated \citep{cardoso-bihloSolverintheloopFrameworkEndtoend2026}.

Recent advances in differentiable programming have introduced a new paradigm for addressing these inverse and optimisation problems by enabling gradients to be propagated through numerical simulations using automatic differentiation (AD) \citep{alhashimControlFlowBehavior2025}. AD systematically evaluates derivatives of computational operations with respect to model inputs, allowing inverse problems to be formulated as gradient-based optimisation rather than relying solely on black-box search or finite-difference approximations \citep{liInundaGPUNativeAgentenabled2026, shenDifferentiableModellingUnify2023}. This approach has recently been extended to hydrodynamic modelling, with differentiable solvers emerging for flood and tsunami applications \citep{cardoso-bihloSolverintheloopFrameworkEndtoend2026, liInundaGPUNativeAgentenabled2026, liWatershedscaleHydrodynamicModeling2026, liuScientificMachineLearning2025, renteriaDifferentiableSolverPhaseresolved2026}. These developments demonstrate the potential of differentiable physics to provide end-to-end frameworks for solving inverse problems while retaining the physical constraints of numerical hydrodynamic models.

In this paper, we present \textbf{DiffSWE2d}, an open-source differentiable SWE solver designed to facilitate inverse and optimisation problems in flood and tsunami modelling. DiffSWE2d solves the shallow water equations using a finite-volume method implemented in PyTorch \citep{paszkePyTorchImperativeStyle}, allowing automatic differentiation to be applied throughout the time-marching simulation. The solver therefore supports both conventional forward modelling and gradient-based optimisation within a unified computational framework. We validate the forward solver using two established benchmark cases and demonstrate its differentiable capability through an application to tsunami waveform inversion. The organisation of this paper is as follows. Section~\ref{sec:2} describes the governing equations, numerical schemes, PyTorch implementation and differentiable framework. Section~\ref{sec:3} presents two benchmark cases for validating the forward solver. Section~\ref{sec:4} demonstrates tsunami waveform inversion using DiffSWE2d. Section~\ref{sec:5} discusses potential applications and limitations and provides conclusions.

\section{Methods}
\label{sec:2}
The core hydrodynamics engine of DiffSWE2d adopts the same governing
equations and numerical schemes used in Basilisk B-Flood \citep{kirstetterBflood10Opensource2021,popinetQuadtreeadaptiveTsunamiModelling2011}
and BG-Flood \citep{bosserelleBGFloodGPUAdaptive2021,xuNearfieldEffects20222025}. 
It solves nonlinear shallow water equations using a second-order
Godunov-type finite volume scheme on a uniform Cartesian grid.

\subsection{Governing equations}
With the hydrostatic pressure assumption, the governing equations are:
\begin{align}
    \frac{\partial h}{\partial t} + \frac{\partial q_{x}}{\partial x} + \frac{\partial q_{y}}{\partial y} &= S_{h} \\
    \frac{\partial q_{x}}{\partial t} + \frac{\partial}{\partial x} \left(\frac{q_{x}^{2}}{h} + \frac{g}{2}h^{2}\right) + \frac{\partial}{\partial y} \left(\frac{q_{x}q_{y}}{h}\right) &= S_{g,x} + S_{f,x} \\
    \frac{\partial q_{y}}{\partial t} + \frac{\partial}{\partial x} \left(\frac{q_{x}q_{y}}{h}\right) + \frac{\partial}{\partial y} \left(\frac{q_{y}^{2}}{h} + \frac{g}{2}h^{2}\right) &= S_{g,y} + S_{f,y}
\end{align}

where $h$ is the water depth, $q_{x}$ and $q_{y}$
are the unit discharges in the $x$ and $y$ directions and $q_{x}=hu$
and $q_{y}=hv$, in which $u$ and $v$ are the depth-averaged velocities in the $x$ an $y$
directions. $S_{h}$ is the mass source and sink term including
rainfall and infiltration. $S_{g,x}$ and $S_{g,y}$
denote the gravitational source terms associated with bed-slope gradients. $S_{f,x}$ and $S_{f,y}$
are friction source terms. The shallow-water equations can be expressed in
conservative form as:
\begin{equation}
    \frac{\partial}{\partial t}U + \frac{\partial}{\partial x}F_{x} + \frac{\partial}{\partial y}F_{y} = S\
\end{equation}

where $U$ is a vector of conserved variables, $F_{x}$ and
$F_{y}$ are the fluxes in the $x$ and $y$ directions defined as

\[ 
U = \begin{bmatrix}
h \\
q_{x} \\
q_{y}
\end{bmatrix}, \quad
F_{x} = \begin{bmatrix}
q_{x} \\
\frac{q_{x}^{2}}{h} + \frac{g}{2}h^{2} \\
\frac{q_{x}q_{y}}{h}
\end{bmatrix}, \quad
F_{y} = \begin{bmatrix}
q_{y} \\
\frac{q_{x}q_{y}}{h} \\
\frac{q_{y}^{2}}{h} + \frac{g}{2}h^{2}
\end{bmatrix} 
\]

The source term $S$ includes mass source term $S_{h}$, gravity
source term $S_{g}$ and friction source term
$S_{f}$.

\[S = S_{h} + S_{g} + S_{f} = \begin{bmatrix}
R \\
0 \\
0
\end{bmatrix} + \begin{bmatrix}
0 \\
 - gh\frac{\partial z_{b}}{\partial x} \\
 - gh\frac{\partial z_{b}}{\partial y}
\end{bmatrix} + \begin{bmatrix}
0 \\
S_{f,x} \\
S_{f,y}
\end{bmatrix}\]

where $R$ is mass source and sink terms including rainfall and infiltration, and
$z_{b}$ is the bed elevation.

\subsection{Time integration}
An explicit second-order predictor-corrector scheme is used for
time-integration \citep{kirstetterBflood10Opensource2021,popinetQuadtreeadaptiveTsunamiModelling2011}. The stable time
step ($\Delta t$) is based on the Courant-Friedrichs-Levy (CFL) condition:

\[\Delta t = \frac{CFL}{\frac{a_{x}}{\Delta x} + \frac{a_{y}}{\Delta y}}\]

where $a_{x}$ and $a_{y}$ are the maximum wave speeds
in the $x$ and $y$ directions. $\Delta x$ and $\Delta y$ are grid sizes.

\subsection{Flux calculation}
Numerical fluxes at cell interfaces are computed using the HLLC
(Harten-Lax-van Leer contact) approximate Riemann solver \citep{toroRestorationContactSurface1994,toroHLLCRiemannSolver2019a}. Spatial reconstruction uses a MUSCL
(\textbf{M}onotonic \textbf{U}pstream-centered \textbf{S}cheme for \textbf{C}onservation \textbf{L}aws) formulation with a minmod slope limiter to achieve second-order spatial accuracy while controlling spurious oscillations near discontinuities. 

To improve the treatment of complex and steep topography, hydrostatic reconstruction is applied following \citet{audusseFastStableWellbalanced2004}, with second-order spatial reconstruction subsequently performed following \citet{buttinger-kreuzhuberNewSecondorderShallow2019,chenNewHydrostaticReconstruction2017}. This treatment addresses limitations associated with hydrostatic reconstruction over steep topographic gradients \citep{delestreLimitationHydrostaticReconstruction2012}.

\subsection{Friction}
Two formulations of bottom friction are available, based either on Manning's roughness coefficient
$n$ or the roughness length $z_{0}$.

If using the Manning's roughness coefficient $n$,

\[
    S_{f,x} = - gn^{2}\frac{q_{x}q}{h^{\frac{7}{3}}}, \quad S_{f,y} = - gn^{2}\frac{q_{y}q}{h^{\frac{7}{3}}}
\]

where \(q = \sqrt{q_{x}^{2} + q_{y}^{2}}\).

If using the roughness length $z_{0}$,
\[
    S_{f,x} = - C_{f}\frac{q_{x}q}{h^{2}}, \quad S_{f,y} = - C_{f}\frac{q_{y}q}{h^{2}}
\]

where $C_{f}$ is the drag coefficient derived from the
logarithmic velocity profile \citep{smartImprovingFloodHazard2018}. The drag coefficient is defined as

\[
C_{f} = \left\{ \begin{array}{r}
\frac{1}{0.46\left( h/z_{0} \right)} \\
\frac{1}{\left\lbrack 2.5\left( \ln\left( \frac{h}{z_{0}} \right) - 1.0 + \frac{1.359}{h/z_{0}} \right) \right\rbrack^{2}}
\end{array} \right. , \quad 
\begin{array}{ll}
\frac{h}{z_{0}} < e & (\text{shallow water}) \\
\frac{h}{z_{0}} \geq e & (\text{deep water})
\end{array}
\]

where e \(\approx\) 2.718. For the New Zealand context, the national
coverage of LiDAR derived roughness length at 8 m resolution is publicly
available \citep{pearsonNationwide8mHydrologically2026} and has been used for national-scale flood mapping \citep{harangFloodHazardAotearoa2026}.

The friction is applied semi-implicitly as an independent update at the
end of each time step after the full hydrodynamic state has been updated \citep{kirstetterBflood10Opensource2021}.

\subsection{PyTorch implementation}
\subsubsection{Model architecture}
The solver (\texttt{SWE2D} class) is structured as a standard PyTorch \texttt{nn.Module}. The variables are stored in a state tensor \textbf{U} of shape \texttt{[batch, 3, ny, nx]}. The three channels represent water depth ($h$), x-directional momentum ($hu$), and y-directional momentum ($hv$). Static fields (e.g., grid size) are registered as non-trainable PyTorch buffers.

\subsubsection{Time stepping}
The \texttt{forward} function is the standard entry point for a PyTorch \texttt{nn.Module}, responsible for advancing the simulation over a simulation duration (\texttt{t\_end}). The \texttt{forward} function orchestrates multiple calls to the \texttt{step} function. The \texttt{step} function executes a single adaptive timestep using a two-stage midpoint predictor-corrector scheme. It calls the \texttt{spatial\_operator} function to calculate the spatial derivatives and updates the state tensor \textbf{U}.

\subsubsection{Spatial operator}
The \texttt{spatial\_operator} function computes the spatial gradients and source terms  using vectorized tensor operations. 

\textbf{Boundary Ghost Cells} The state tensor is padded using \texttt{add\_ghost\_cells} to handle reflective, transmissive, or other boundaries seamlessly within the tensor convolution limits.

\textbf{Spatial Reconstruction} Interface states (Left/Right) are interpolated using \texttt{bflood\_cn\_reconstruction}. This step includes the slope limiters to maintain high-order accuracy while preventing spurious oscillations near shocks.

\textbf{HLLC Riemann Solver} The intercell numerical fluxes ($F_x$, $F_y$) and local maximum wave speeds are computed in \texttt{bflood\_hllc\_flux}. Because PyTorch supports vectorized conditional logic (\texttt{torch.where}), the solver evaluates all possible Riemann wave regions simultaneously across the entire grid without iterative loops.

\subsubsection{Differentiability}
The differentiable solver leverages the automatic differentiation (AD) framework in PyTorch to enable gradient propagation through the time-marching simulation. The solver supports both a forward pass, in which the SWE are integrated in time, and a backward pass, in which gradients of an objective function are propagated through the numerical solver. During the forward pass, the sequence of differentiable mathematical operations is recorded as a computational graph. During the backward pass, PyTorch applies reverse-mode automatic differentiation to this computational graph to evaluate the gradient of the loss function with respect to the model inputs and parameters, including initial conditions, boundary conditions, and other physical parameters \citep{cardoso-bihloSolverintheloopFrameworkEndtoend2026}. The trainable parameters are enabled using \texttt{torch.nn.Parameter}.

Several modifications to the numerical solver are required to improve differentiability in the presence of discontinuities inherent in shallow water flow. In particular, a continuous water film is introduced at wet--dry interfaces to avoid abrupt changes in the computational state that can lead to undefined or unstable gradients. For limiter choices and branch selection, PyTorch's native automatic differentiation automatically records the active branch during the forward pass. During the backward pass, gradients are routed only through the active branch. In addition, the adaptive CFL time step is detached from the computational graph during backpropagation. The adaptive time step is therefore retained during the forward simulation to satisfy the CFL stability condition, while its dependence on the model state is ignored when computing gradients. This treatment prevents gradients from propagating through the discrete time-step selection and improves the robustness of gradient backpropagation.

\begin{figure}[!ht]
    \centering
    \includegraphics[width=0.4\textwidth]{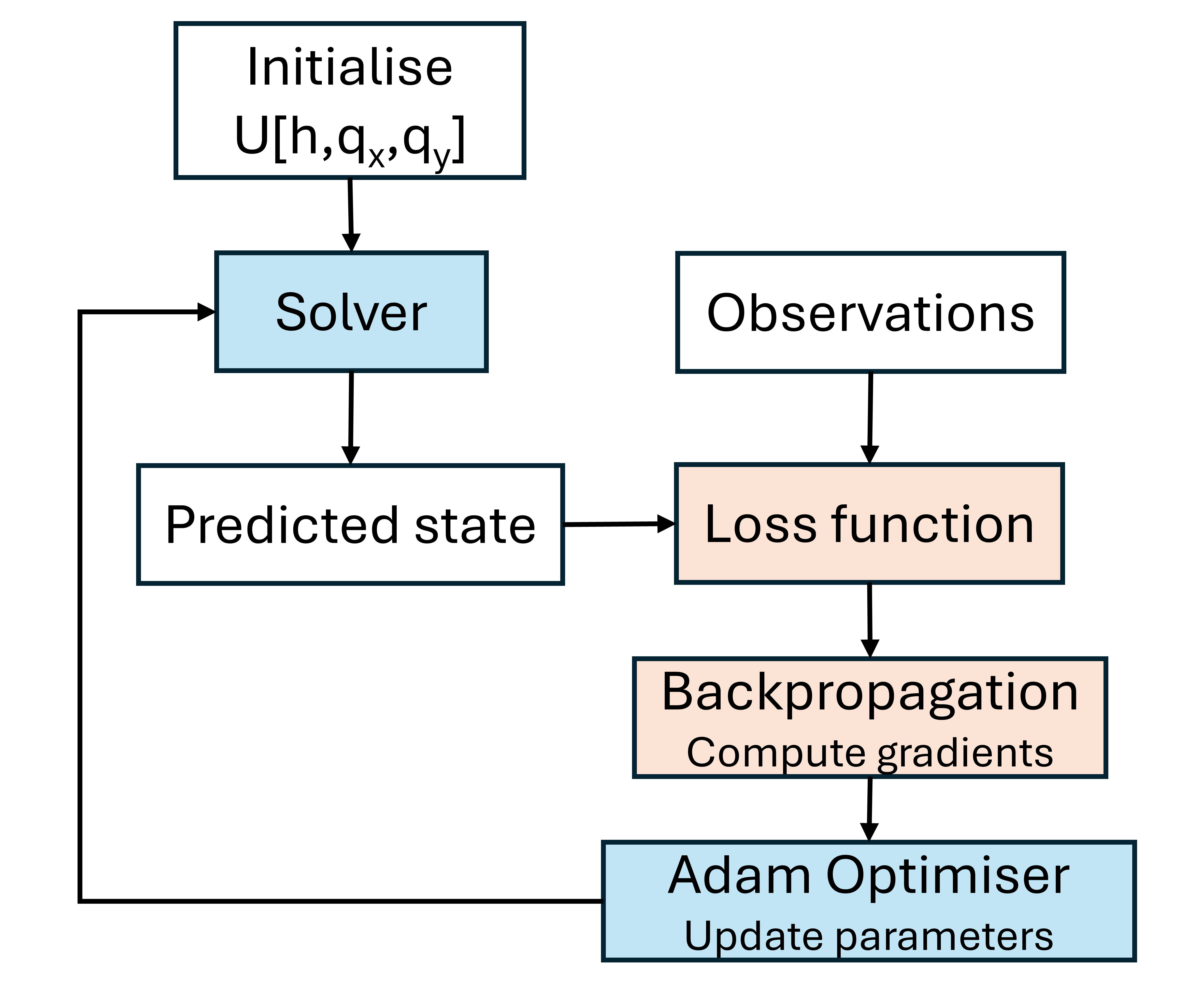}
    \caption{Flowchart of solving an inverse problem using the differentiable solver. Blue boxes represent processes in the numerical solver. Yellow boxes represent processes using automatic differentiation.}
    \label{fig:fig1}
\end{figure}

Figure~\ref{fig:fig1} illustrates the workflow for solving an inverse problem using the differentiable solver. Given a set of initial model parameters, the solver first performs a forward simulation and computes the model outputs. These outputs are then compared with observations or target values through a loss function. The backward pass propagates the resulting gradients through the computational graph to the model inputs, which can subsequently be updated using a gradient-based optimisation algorithm.

\subsubsection{Computational acceleration}
The solver supports the \texttt{torch.compile()} functionality introduced in PyTorch 2.0. This just-in-time (JIT) compilation mechanism transforms PyTorch operations into optimised kernels at runtime, which can reduce computational overhead and accelerate model execution, particularly for large-scale simulations.

\subsubsection{File format}
The solver supports digital elevation models (DEMs) provided in either ASCII or NetCDF format. Bottom roughness can be specified as either a spatially uniform constant or a spatially varying field. Rainfall forcing can similarly be prescribed as a constant rate or as spatially and temporally varying rainfall fields.

\section{Validation of the forward modelling}
\label{sec:3}
In this section we validate the solver with two benchmark tests with
forward modelling.

\subsection{Pluvial flooding}
\label{sec:pluvial}
\begin{figure}[!ht]
    \centering
    \includegraphics[width=0.5\textwidth]{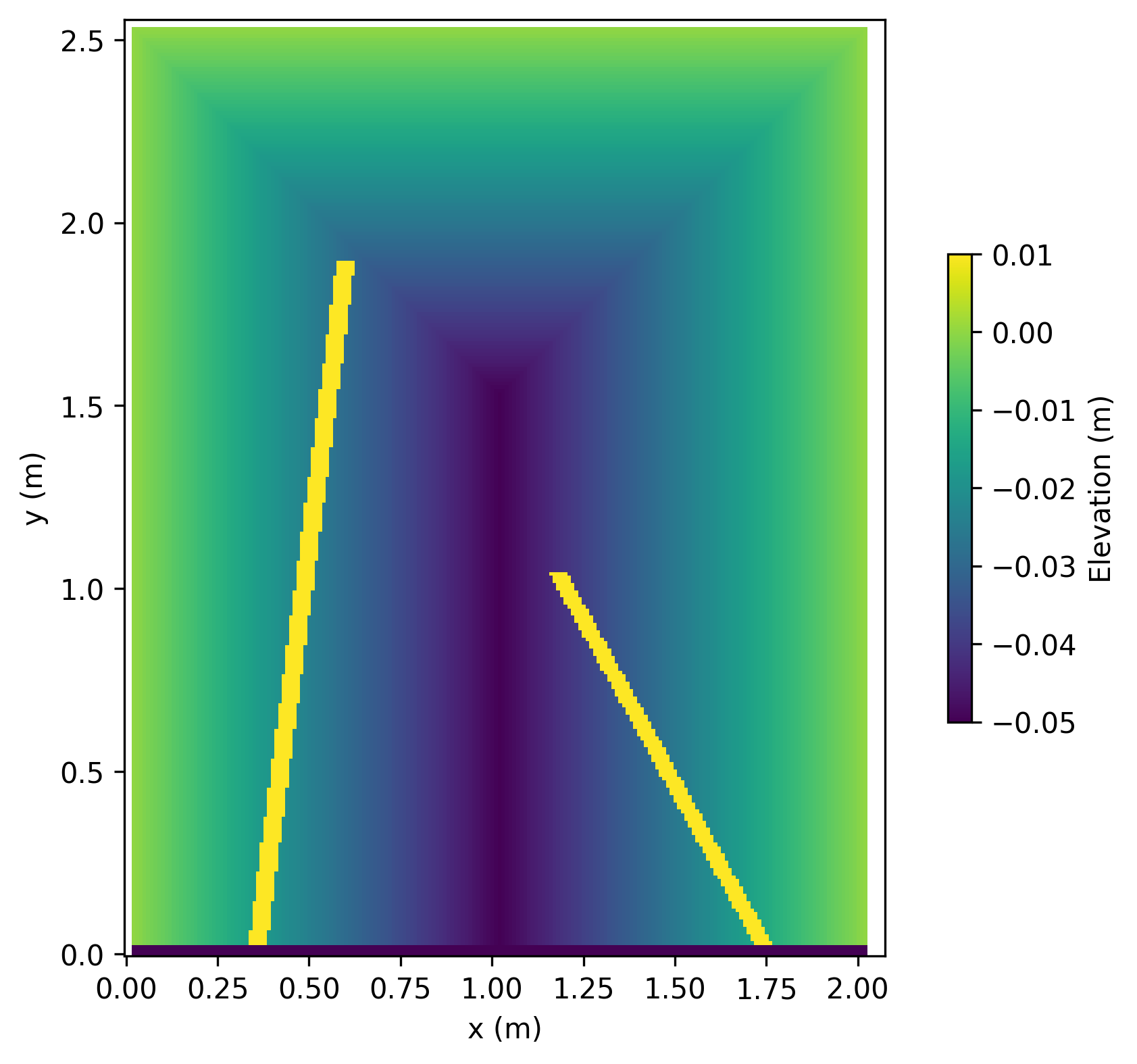}
    \caption{Topography of the experimental basin.}
    \label{fig:fig2}
\end{figure}

\begin{figure}[!ht]
    \centering
    \includegraphics[width=0.9\textwidth]{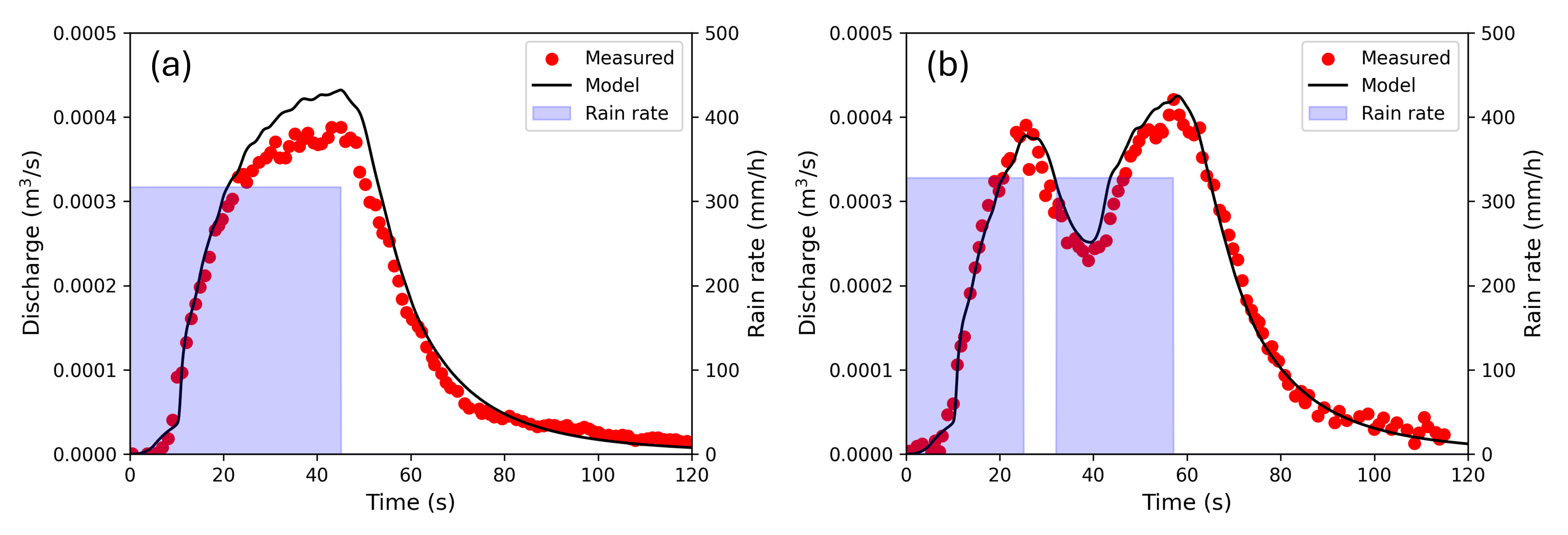}
    \caption{Modelled outlet discharges compared with measured discharges for
rainfall runoff experiments (a) C1 and (b) C3.}
    \label{fig:fig3}
\end{figure}

The rainfall-runoff experiments in \citet{ceaHYDROLOGICFORECASTINGFAST} were simulated to evaluate the capability of DiffSWE2d to model runoff generation and discharge response to precipitation forcing. The experimental setup consists of a rectangular basin bounded by two walls, which extends the flow path and consequently increases the duration of the outlet hydrograph (Figure~\ref{fig:fig2}). The model setup
includes a uniform grid size of 0.01 m and a Manning's roughness
coefficient $n$ = 0.012. The flow undergoes a free fall before exiting the
outlet boundary. The simulated outlet discharge agrees well with the recorded hydrograph, reproducing the overall timing and magnitude of the observed discharge response (Figure~\ref{fig:fig3}).

\subsection{Monai Valley tsunami benchmark}
\label{sec:3.2}
\begin{figure}[!ht]
    \centering
    \includegraphics[width=0.6\textwidth]{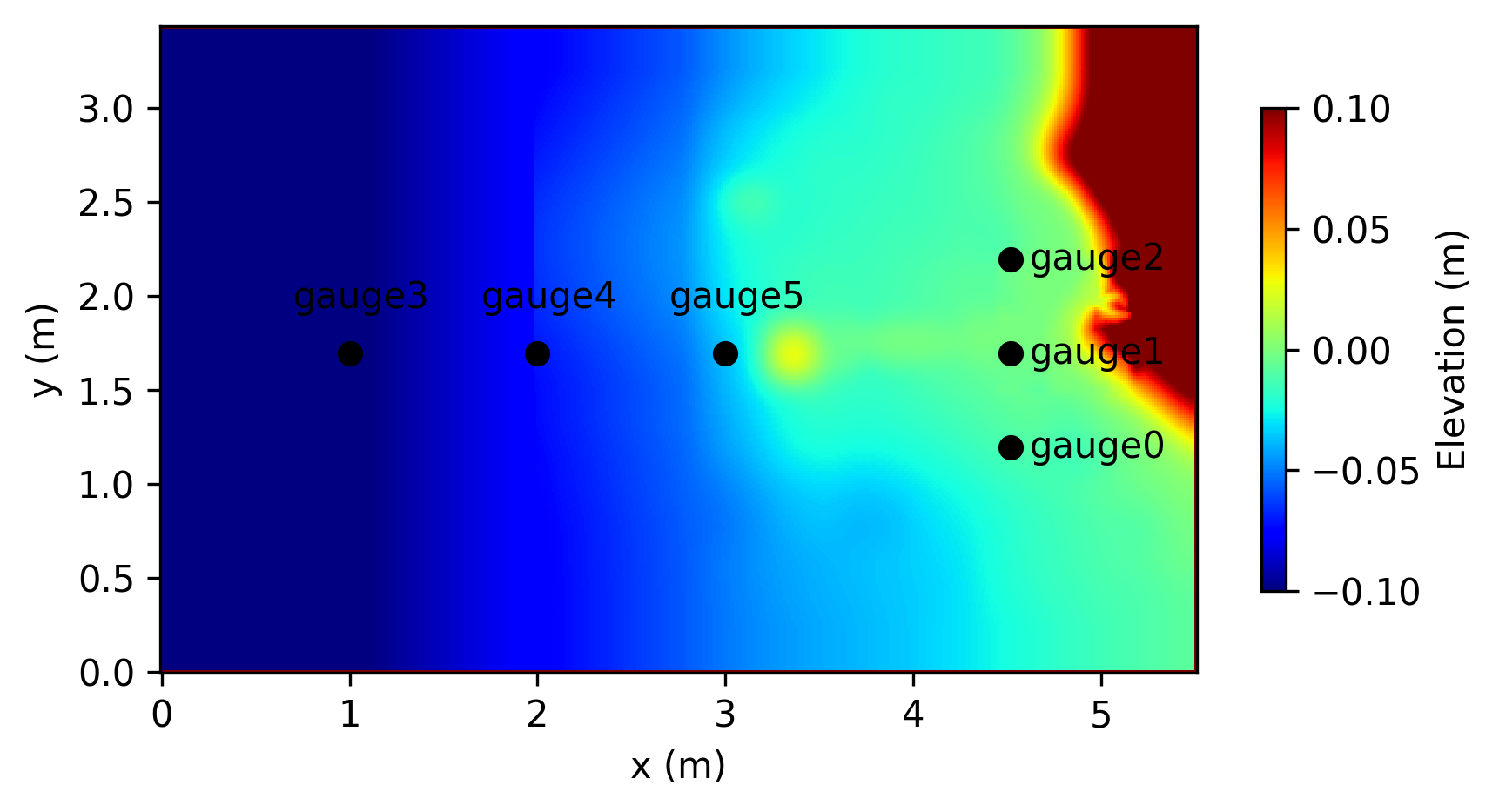}
    \caption{Bathymetry of the Monai Valley test with gauge locations. Gauges 0-2 are experimental gauges where water levels were recorded. Gauges 3-5 are hypothetical gauges referenced in Section~\ref{sec:4}. Incident waves are from the left boundary.}
    \label{fig:fig4}
\end{figure}

\begin{figure}[!ht]
    \centering
    \includegraphics[width=0.6\textwidth]{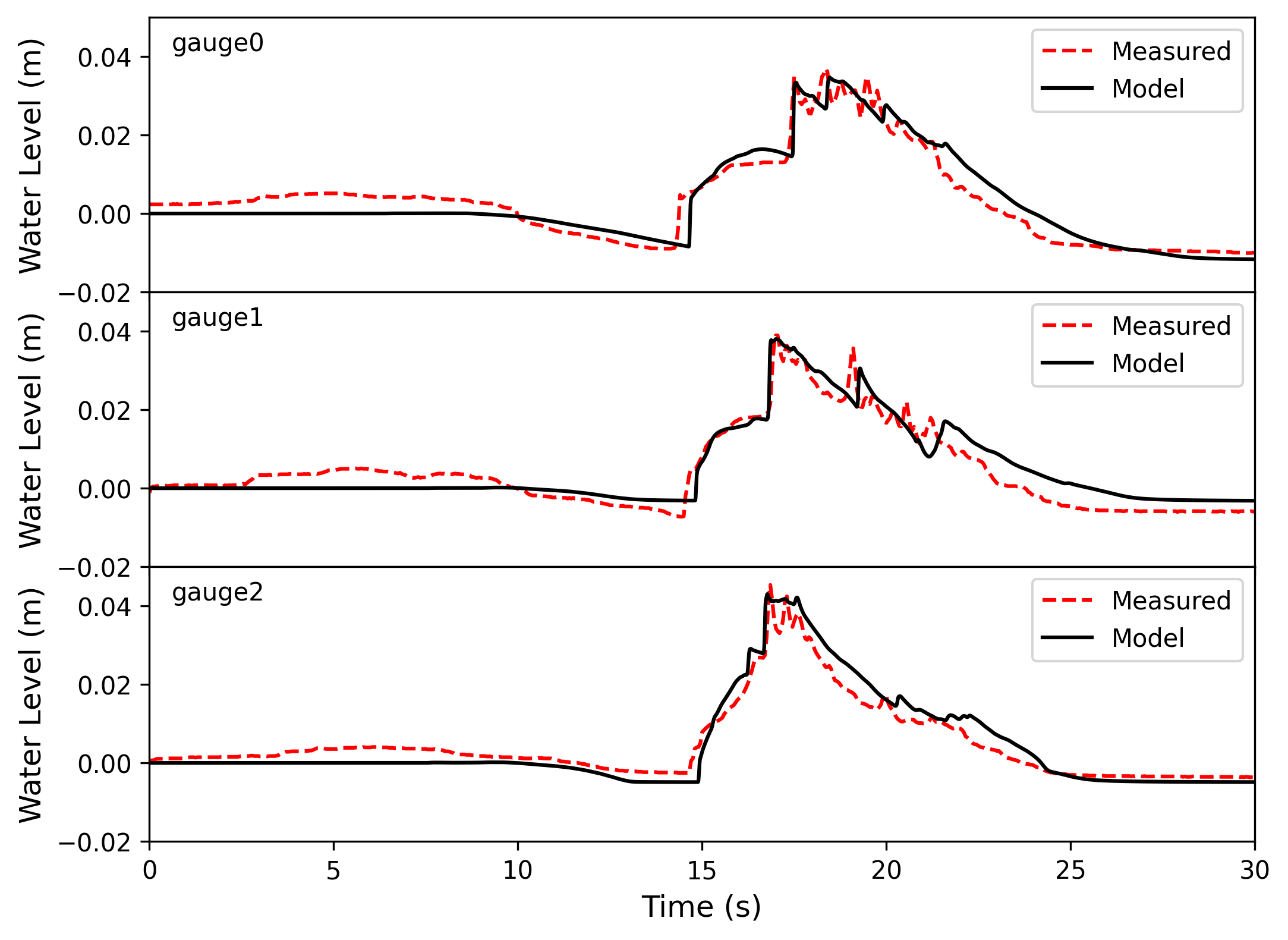}
    \caption{Performance of DiffSWE2d on the Monai Valley benchmark. The
dashed red lines show the measured water levels and the black lines show the
simulated results.}
    \label{fig:fig5}
\end{figure}

The Monai Valley tsunami benchmark is an experimental set up of the
Monai Valley in Okushiri Island, Japan, where the 1993
Hokkaido-Nansei-Oki tsunami resulted in extreme wave runup in the narrow
and steep coastal inlet (Figure~\ref{fig:fig4}). A uniform spatial resolution of 0.01 m was used
in the model setup. The Manning's $n$ = 0.01 was used as the friction
coefficient. Three gauges (gauges 0-2) recorded water levels at the
measurement locations in the experiment \citep{synolakisValidationVerificationTsunami2008}. DiffSWE2d reproduces the
wave accurately at the three gauges and the results are similar to those
using BG-Flood \citep{bosserelleBGFloodGPUAdaptive2021} and Basilisk \citep{popinetQuadtreeadaptiveTsunamiModelling2011} as shown in Figure~\ref{fig:fig5}.

\section{Tsunami waveform inversion from observations: an example}
\label{sec:4}
Existing approaches for tsunami source inversion use Green's function or adjoint methods \citep{satake1987inversion,tsushima2012tsunami,pires2001tsunami}. In this section, we demonstrate an alternative approach based on automatic differentiation, in which DiffSWE2d directly propagates gradients from observed tsunami waveforms to the prescribed boundary waveform. The objective is to recover the incident tsunami waveform from water-level observations at a limited number of gauge locations. The Monai Valley tsunami benchmark in Section~\ref{sec:3.2} was used as an example.

\subsection{Model}
The incident tsunami waveform at the model boundary was parameterised using a clamped cubic B-spline with 20 control points. The control points were treated as trainable parameters and optimised using gradient-based optimisation. The inversion was initialised with a zero water-level perturbation at the boundary. DiffSWE2d was then used to perform a forward simulation, and the simulated water levels at the observation gauges were compared with the target gauge records. The resulting objective function was differentiated through the complete time-marching simulation, and the gradients with respect to the B-spline control points were used to update the boundary waveform using the Adam optimiser. 

\subsubsection{Workflow for optimising the boundary waveform}
The detailed workflow for optimising the boundary waveform is shown in the five steps below.

1. Initialise or optimize a vector of 20 control values \(raw_j\). 

2. Convert each raw control to a bounded physical value constrained by the allowed range using a sigmoid function. For each control \(c_j\),
\[c_j = c_{\min} + \left(c_{\max} - c_{\min}\right)\sigma\left(\mathrm{raw}_j\right)\]

where \(c_{min}\) and \(c_{max}\) are the predefined value range.

3. Build a B-spline basis matrix and reconstruct the full boundary series \(b(t_i)\). 
\[b(t_i) = \sum_{j=1}^{m} B_j(t_i) c_j\]

where \(B_j(t_i)\) are the spline basis values at time \(t_i\) used as weights and \(m\) = 20. The B-spline basis function was built using the Cox-de Boor recursion with a degree of 3. The spline basis values sum up to 1 at every time point, which keeps the spline bounded and smooth.

4. Feed boundary series into the forward pass. The computation graph is,

\[\text{raw controls} \rightarrow c \rightarrow b \rightarrow \text{simulation} \rightarrow L\]

5. Backpropagate the loss \(L\) through the computation graph back to the control parameters using \texttt{backward()}. The gradient flows as:

\[\frac{\partial L}{\partial \mathrm{raw}_j} = \frac{\partial L}{\partial b} \cdot \frac{\partial b}{\partial c_j} \cdot \frac{\partial c_j}{\partial \mathrm{raw}_j}\]

where \[\frac{\partial c_j}{\partial \mathrm{raw}_j} = (c_{\max} - c_{\min}) \sigma(\mathrm{raw}_j) (1 - \sigma(\mathrm{raw}_j))\] and \[\frac{\partial b}{\partial c} = B\]

In this way, the gradient with respect to the raw controls is effectively the spline-basis-weighted sensitivity of the loss.

\subsubsection{Model set-up}
The initial learning rate was set to 0.02 and subsequently reduced using a learning-rate scheduler, with a minimum learning rate of \(1.0 \times 10^{- 5}\). Optimisation was terminated when the objective function failed to improve by more than \(1.0 \times 10^{- 8}\) for 10 consecutive epochs.

The total loss function consists of the mean squared error (MSE) between the simulated and target water levels at the training gauges (training loss), together with three regularisation terms applied to the reconstructed boundary waveform:

\[L = L_{train} + \lambda_{1}L_{d1} + \lambda_{2}L_{d2} + \lambda_{3}L_{amp}\]

where \(L_{train}\) is the average MSE across the training gauges; The first- and second-order smoothness terms are defined as
\begin{align*}
    L_{d1} &= \mathrm{mean}\left( {(b_{t + 1} - b_{t})}^{2} \right) \\
    L_{d2} &= \mathrm{mean}\left( {(b_{t + 2} - 2b_{t + 1} + b_{t})}^{2} \right)
\end{align*}
respectively. The amplitude regularisation term is defined as
\begin{equation*}
    L_{amp} = mean\left( {b_{t}}^{2} \right)
\end{equation*}
where \(b_{t}\), \(b_{t + 1}\) and \(b_{t + 2}\) denote consecutive values of the reconstructed boundary waveform. The corresponding regularisation weights are
\(\lambda_{1}=1.0\times10^{-3}\), \(\lambda_{2}=1.0\times10^{-2}\), and \(\lambda_{3}=1.0\times10^{-6}\).

\begin{figure}[!ht]
    \centering
    \includegraphics[width=0.6\textwidth]{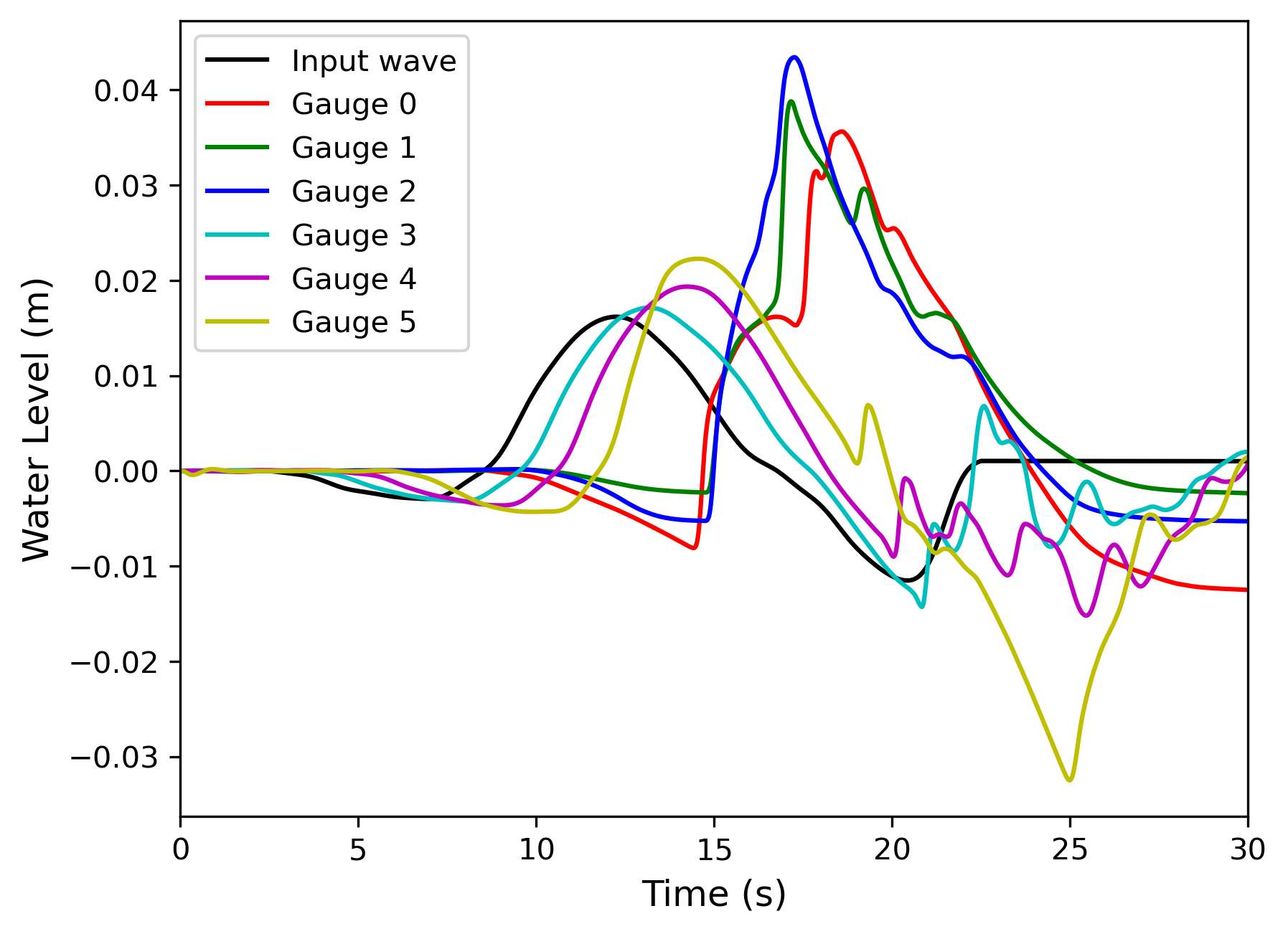}
    \caption{Input waves and simulated waves at gauges 0-5.}
    \label{fig:wave_gages_data}
\end{figure}

Two inversion scenarios were considered. In Scenario (a), observations from gauges 0 and 1 were used as training data, while observations from gauge 2 were withheld for independent evaluation. In Scenario (b), water-level data from gauges 0, 1, 3, 4, and 5 were used for training, with gauge 2 again reserved for testing. Scenario (b) was designed to investigate the effect of increasing the number of observation gauges on the accuracy of waveform reconstruction. For this scenario, the measured incident waveform was first used as the boundary forcing in a forward simulation to generate water-level data at gauges 0-5 (Figure~\ref{fig:fig4}). The resulting synthetic observations had a spatial resolution of 0.08~m. Figure~\ref{fig:wave_gages_data} shows the input waves and the simulated waves at gauges 0-5.

\subsection{Results}
\begin{figure}[!ht]
    \centering
    \includegraphics[width=0.8\textwidth]{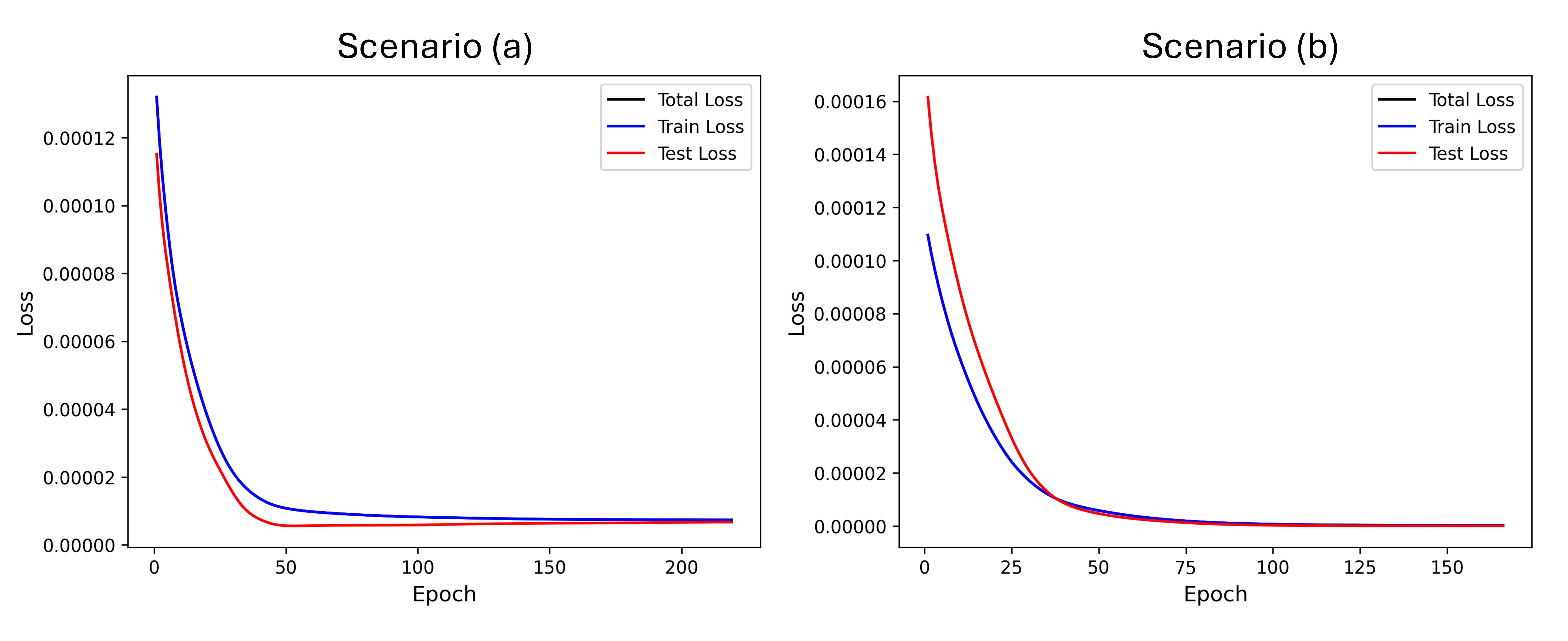}
    \caption{Evolution of total loss, training loss and testing loss.}
    \label{fig:fig6}
\end{figure}

\begin{figure}[!ht]
    \centering
    \includegraphics[width=0.8\textwidth]{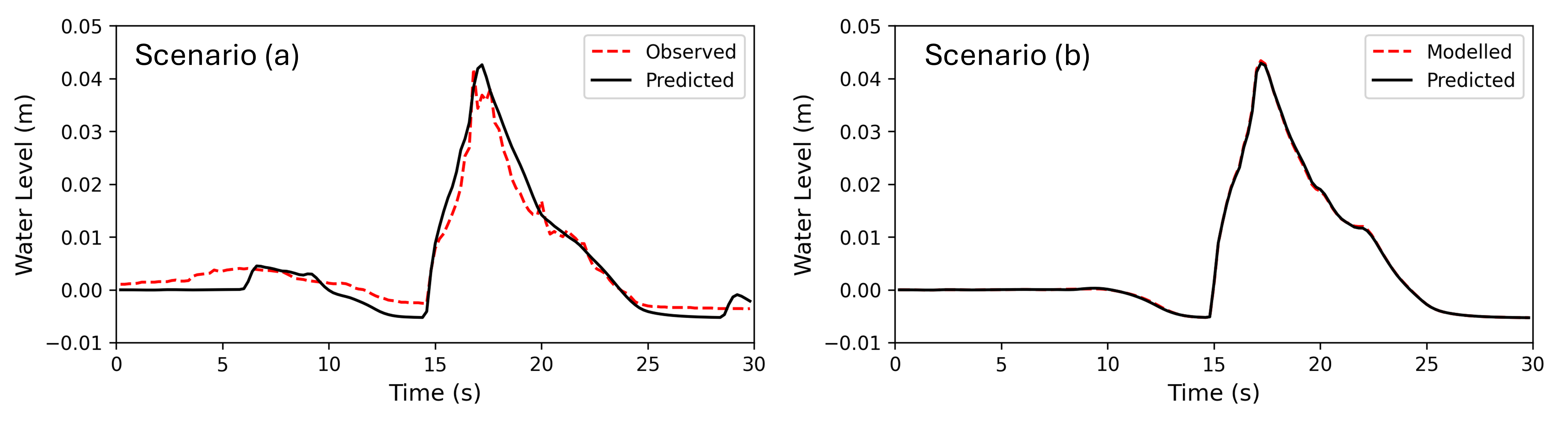}
    \caption{Comparison of original test gauge water levels (observed or modelled) 
    with prediction from inferred input waves.}
    \label{fig:fig7}
\end{figure}

\begin{figure}[!ht]
    \centering
    \includegraphics[width=0.8\textwidth]{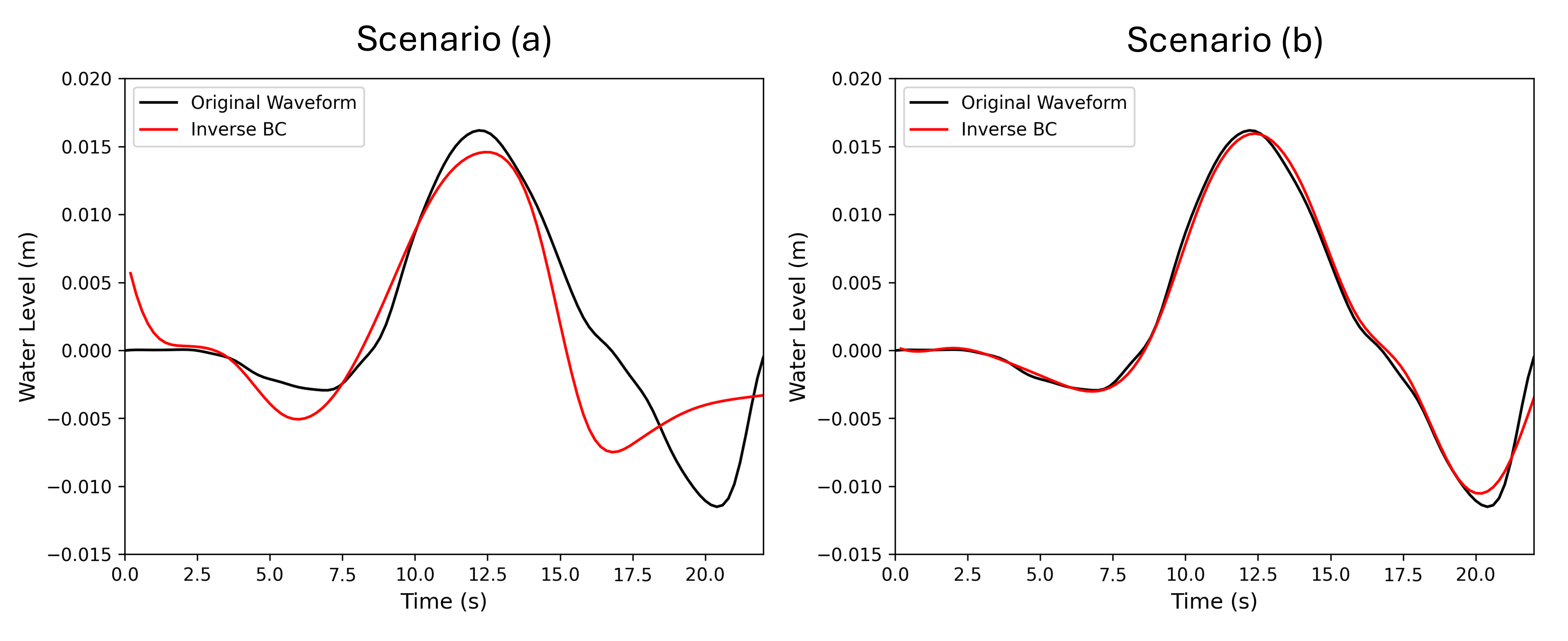}
    \caption{Comparison of raw input waves with inferred input waves.}
    \label{fig:fig8}
\end{figure}

Scenario (a) stopped training after 219 epochs and Scenario (b) stopped after 166 epochs. Figure~\ref{fig:fig6} shows the evolution of the total loss, training loss, and testing loss during the inversion. In both scenarios, the total loss closely follows the training loss, indicating that the contribution of the boundary regularisation terms is small relative to the data-misfit term. Both scenarios reproduce the water-level evolution at the withheld testing gauge (gauge 2) reasonably well, as shown in Figure~\ref{fig:fig7}. The reconstructed waveforms at the model boundary provide further insight into the effect of the number of observation gauges (Figure~\ref{fig:fig8}). In Scenario (a), using observations from only two gauges allows the general characteristics of the incident waveform to be recovered, but the reconstructed waveform underestimates both the positive peak and negative trough and exhibits a distorted wave shape. In contrast, Scenario (b), which uses observations from five training gauges, produces a substantially closer match to the reference boundary waveform. These results demonstrate that additional spatially distributed observations can provide stronger constraints on the inverse problem and improve the reconstruction of the incident tsunami waveform.

\section{Conclusion}
\label{sec:5}
In this paper, we introduced DiffSWE2d, an open-source differentiable shallow water equations solver for end-to-end flood and tsunami modelling implemented in PyTorch. By leveraging automatic differentiation, DiffSWE2d can not only perform forward modelling, as in traditional solvers, but also backpropagate gradients with respect to input parameters, enabling gradient-based optimisation and inverse modelling. This capability provides an efficient framework for solving inverse and optimisation problems, including tsunami waveform inference, bathymetry and roughness estimation, topographic optimisation for flood mitigation, and sensor deployment optimisation.

Although the solver is differentiable with respect to key input parameters, including DEM, roughness and boundary conditions, the computational graph is not fully differentiable through all components of the numerical scheme. In particular, adaptive time-step selection, wet/dry transitions, limiter choices, flux-region selection, and gradient clipping introduce non-differentiable or detached operations. Furthermore, backpropagation through the computational graph can be computationally expensive, particularly for long simulation sequences, and shares some of the challenges associated with training Recurrent Neural Networks (RNNs), such as increasing memory requirements with sequence length. Future development will focus on coupling the differentiable solver with neural networks to accelerate the inversion process and enable more efficient end-to-end learning and optimisation for flood and tsunami modelling.

\section*{Code and data availability statement}
The code and examples are publicly available at:
\url{https://github.com/ZhonghouXu/DiffSWE2d}

\section*{Acknowledgements}

This research was supported by the MBIE Smart Ideas project:
AI-enhanced compound flood model for real-time extreme hazard forecasts
(NIW2473) and Strategic Science Investment Fund (SSIFTC2701). The author thanks Cyprien Bosserelle for providing the data
for the benchmarking case in Section~\ref{sec:pluvial} and reviewing the initial manuscript.

\bibliographystyle{abbrvnat}  
\bibliography{references}  


\end{document}